\documentclass[prl,twocolumn,showpacs,amsmath,amssymb]{revtex4}
\usepackage{graphicx}
\renewcommand{\v}[1]{{\bf #1}}

\newcommand{\w}{{\omega}}

\def\eqa{\begin{eqnarray}}
\def\eea{\end{eqnarray}}
\newcommand{\eq}{\begin{equation}}
\newcommand{\ee}{\end{equation}}

\renewcommand{\>}{\rangle}

\newcommand{\ua}{\uparrow}
\newcommand{\da}{\downarrow}
\newcommand{\ra}{\rightarrow}

\newcommand{\Ga}{\Gamma}

\begin{document}

\title{Two-mode variational Monte Carlo study of Quasiparticle excitations in cuprates}
\author{Fei Tan and Qiang-Hua Wang}
\address{National Laboratory of Solid State Microstructures \& Department of Physics, Nanjing University,
Nanjing 210093, China}


\begin{abstract}
Recent measurements of quasiparticles in hole-doped cuprates
reveal highly unusual features: 1) the doping-independent Fermi
velocity, 2) two energy scales in the quasiparticle spectral
function, and 3) a suppression of the low energy spectral weight
near the zone center. The underlying mechanism is under hot
debate. We addressed these important issues by a novel two-mode
variational Monte Carlo (VMC) study of the t-J model. We obtained
results in agreement with the experiments but without invoking
extrinsic effects. Besides, we resolved a long standing issue of
the sum rule for quasiparticle spectral weights in VMC studies.
The electron doped case was also discussed.
\end{abstract}

 \maketitle

Understanding the quasiparticle excitations is a preclude to
unravel the mechanism of high-Tc superconductivity. Recent
angle-resolved photo-emission spectra (ARPES) in cuprates reveal
highly unusual features and lead to hot debates. For example, on
the low energy quasiparticle band the nodal Fermi velocity does
not seem to increase with increasing doping, and is called a
"universal nodal Fermi velocity".\cite{zhou} This seems to
challenge the concept of doped Mott insulators which would naively
predict a vanishing Fermi velocity with decreasing doping. On the
other hand, ARPES measurements up to higher energies
\cite{feng,lanzara,ding,Meevasana,valla,Chang,Inosov,undoped}
revealed that the quasi-particle dispersion along the nodal
direction breaks up near the momentum ($\pi/4$, $\pi/4$) at an
energy around $0.3\sim 0.4$ eV (or the low energy spectral weight
near the $\Ga$ seems to disappear), and reappeared around 1eV. The
two bands are connected in a waterfall fashion in the
momentum-energy space. This high energy anomaly has intrigued many
theoretical studies and
disputes.\cite{feng,Phillips,twobandMF,DCA,Byczuk,exactD,Markiewicz,zhouT,Greco}
A central concern is whether the anomaly is caused by extrinsic
effects due to phonons, or is an intrinsic property of the
strongly correlated electrons themselves. We summarize the main
important issues that will concern us in this Letter: 1) the
doping-independent Fermi velocity, 2) two energy scales in the
quasiparticle spectral function, and 3) a suppression of the low
energy spectral weight near the zone center.

We address the above important issues within the commonly accepted
one-band t-J model without invoking other extrinsic effects. The
theoretical machinery we use is the Gutzwiller projection
variational Monte Carlo (VMC). It proves to give good energy and
supports d-wave pairing
symmetry.\cite{GrosRVB,Paramekanti,Yokoyama,Ivanov,LiTK} As for
quasiparticle excitations under concern, a projected mean-field
excited state is commonly used in the
literature.\cite{Yunoki,Bieri,LiT} For a given momentum $\v k$ and
spin $\sigma$, only one such quasiparticle can be constructed. We
call such an approach as a single-mode approach (SMA).
Unfortunately, the SMA gives only a single low energy band and
predicts that the spectral weight is maximal at the center of the
Brillouine zone (the $\Gamma$ point), in contrast to the two
energy scales and suppression of low energy spectral weight
revealed by ARPES. The spectral weight below the Fermi level
captured by SMA is proved to be smaller than the exact value
$1-x$.\cite{Bieri,sumrule} It is therefore likely that the high
energy feature is due to the missing spectral weight beyond the
SMA. Moreover, the nodal Fermi velocity increases significantly
with doping in such an approach.\cite{Yunoki} In this Letter we
design a two-mode approach (TMA) for each set of quasiparticle
quantum number $(\v k,\sigma$). We show that the TMA satisfies the
spectral sum rules for quasiparticle excitations both below and
above the Fermi level, resolving a long standing problem
SMA.\cite{Edegger} In this way spectral weights beyond the scope
of the SMA can also be captured. Our numerical results are in
agreement with the main experimental features listed above. We
also demonstrate that in the electron doped case there should be
no high energy anomaly below the Fermi level, in agreement with
the finite temperature exact diagonalization.\cite{exactD}

The t-J model Hamiltonian is $H=H_t+H_J$, with
$H_t=-\sum_{\<ij\>,\sigma} t_{ij}(P_G c_{i\sigma}^\dagger
c_{j,\sigma}P_G+{\rm h.c.})$ and $H_J=J\sum_{\<ij\>} (S_i\cdot
S_j-\frac{1}{4}P_G n_i n_j P_G)$.\cite{zrs} Here $t_{ij}=t_1, t_2,
t_3$ are hopping integral between the nearest, second-nearest and
third-nearest neighbor sites $i$ and $j$, $c_{i\sigma}$ is the
electron annihilation operator, $S_i$ is the electron spin,
$n_i=n_{i\ua}+n_{i\da}$ with $n_{i\sigma}=c_{i\sigma}^\dagger
c_{i\sigma}$, and $P_G=\Pi_i(1-n_{i\ua}n_{i\da})$ is the
Gutzwiller projection operator that removes any double
occupations. A widely used trial ground state with d-wave pairing
is $ |\Psi_{GS}\rangle=P_G
P_{N}|\Psi_{dBCS}\rangle$,\cite{GrosRVB} where $P_{N}$ is the
projection operator that fixes the number of electrons,
$|\Psi_{dBCS}\rangle=\prod_{k}(u_k+v_{k}c_{k\uparrow}^{\dagger}c_{-k\downarrow}^{\dagger})|0\rangle$.
Here $u_{k}^{2}=\frac{1}{2}(1+\frac{\xi_{k}}{E_{k}})$,
$v_{k}^{2}=\frac{1}{2}(1-\frac{\xi_{k}}{E_{k}})$,
$E_{k}=\sqrt{\xi_{k}^{2}+\Delta_{k}^{2}}$, $\xi_{k}=-2(\cos
k_x+\cos k_y)-4t_{var}^{'}\cos k_x\cos k_y-2t_{var}^{''}(\cos
2k_x+\cos 2k_y)-\mu_{var}$, $\Delta_k=\Delta_{var}(\cos k_x-\cos
k_y)$. $t_{var}^{'}$, $t_{var}^{''}$, $\Delta_{var}$ and
$\mu_{var}$ are variational parameters. In the following we
concentrate on the quasiparticle excitations. Candidates for such
excitations are described by the quasi-hole, bare-hole,
quasi-electron and bare-electron wave functions as follows: \eqa
|\Psi_{qh}(k,\sigma)\rangle &=& P_{G}P_{N-1}\gamma_{k\sigma}^{\dagger}|\Psi_{dBCS}\rangle,\label{qh}\\
|\Psi_{bh}(k,\sigma)\rangle &=& c_{-k,-\sigma}|\Psi_{GS}\rangle,\label{bh}\\
|\Psi_{qe}(k,\sigma)\rangle &=&
P_{G}P_{N+1}\gamma_{k\sigma}^{\dagger}|\Psi_{dBCS}\rangle,\label{qe}\\
|\Psi_{be}(k,\sigma)\rangle &=&
c_{k,\sigma}^{\dagger}|\Psi_{GS}\rangle.\label{be} \eea Here
$\gamma_{k\sigma}=u_{k}c_{k\sigma}-\sigma
v_{k}c_{-k\bar{\sigma}}^{\dagger}$ is the Bogoliubov
quasi-particle annihilation operator with momentum $k$ and spin
$\sigma$. As pointed out by Ran et al, all these wave functions on
for a specific real-space electron configuration can be written in
the form of determinants.\cite{RanY} Thus the overlaps between
these states can be computed statistically. To ease further
discussion we denote the normalized wave functions for the ground
state, the quasi-hole, the bare-hole, the quasi-electron and the
bare-electron states as, respectively, $|GS\rangle$, $|QH\rangle$,
$|BH\rangle$, $|QE\rangle$ and $|BE\rangle$. The subscripts $k$
and $\sigma$ are left implicit.

The ARPES experiment measures the one particle spectral function
$A(k,\omega)$ in the occupied side ($\w<0)$. According to the
Lehnmann representation,
\begin{align} A(k,\omega)=&\sum_{n,\sigma}[|\langle
n|c_{k\sigma}^{\dagger}|0\rangle|^{2}\delta(\omega+\omega_0-\omega_{n})\nonumber\\&+|\langle
n|c_{-k,-\sigma}|0\rangle|^2\delta(\omega-\omega_0+\omega_{n})].\end{align}
Here $|n\>$ denotes an excited eigenstate of $H-\mu N$ with the
eigenvalue $\w_n$. We define $Z_n^-=|\<n|c_{k\sigma}|0\>|^2$ and
$Z_n^+=|\<n|c_{k\sigma}^\dagger |0\>|^2$ as the spectral weights
of $|n\>$ in the occupied and unoccupied sides, respectively. For
a free particle system, the state with the quantum numbers $(\v
k,\sigma)$ is unique. This is no longer the case in an interacting
system. We recall that $|QH\rangle$ is the only state used in SMA.
As mentioned above, the spectral weights not captured by SMA are
likely to appear at higher energies, and this motivates us to go
beyond the SMA by enforcing the sum rule.

In principle one should construct a complete set of excitations to
satisfy the sum rule. This is possible by exact diagonalization
but is limited by the lattice size. We therefore take a simpler
route but still keep the sum rule. The idea is as follows. Since
$[c_{k\sigma},P_G]\neq 0$, $|QH\>$ and $|BH\>$ are neither
identical nor orthogonal. We can reconstruct two orthogonal states
$|BH\>$ and
$|\eta\>=\frac{1}{\sqrt{1-|\kappa|^2}}(|QH\>-\kappa|BH\>)$ for
each set of quantum numbers $(\v k,\sigma)$, where $\kappa=\langle
BH|QH\rangle$. The Hamiltonian matrix in this basis space is \eqa
(H)_{2\times 2} =\left(
    \begin{array}{cc}
      \langle \eta|H|\eta\rangle & \langle\eta|H|BH\rangle \\
      \langle BH|H|\eta\rangle & \langle BH|H|BH\rangle
    \end{array}\right),\eea which we diaganolize to get the eigenstates $|1\>$ and $|2\>$
as linear combinations of $|BH\>$ and $|\eta\>$ (and thus of
$|BH\>$ and $|QH\>$. The occupied spectral weight for $(\v
k,\sigma)$ is now given by \begin{align}
&Z_{k\sigma,1}^{-}+Z_{k\sigma,2}^{-}\nonumber\\&=\frac{|\langle
1|c_{-k,-\sigma}|\Psi_{GS}\rangle|^2}{\langle\Psi_{GS}|\Psi_{GS}\rangle}+\frac{|\langle
2|c_{-k,-\sigma}|\Psi_{GS}\rangle|^2}{\langle\Psi_{GS}|\Psi_{GS}\rangle}\nonumber\\&
=\frac{\langle \Psi_{bh}|(|1\rangle\langle 1|+|2\rangle\langle
2|)|\Psi_{bh}\rangle}{\langle
\Psi_{bh}|\Psi_{bh}\rangle}\frac{\langle
\Psi_{bh}|\Psi_{bh}\rangle}{\langle\Psi_{GS}|\Psi_{GS}\rangle}\nonumber\\&
=\frac{\langle
\Psi_{bh}|\Psi_{bh}\rangle}{\langle\Psi_{GS}|\Psi_{GS}\rangle}=n_{-k,-\sigma},\end{align}
where we used definition of the unnormalized bare-hole state. We
also used the crucial fact that $|\Psi_{bh}\>$ lies within the
space spanned by $|1\>$ and $|2\>$, or equivalently by $|BH\>$ and
$|\eta\>$. By summing over $(\v k,\sigma)$ we get a total occupied
spectral weight $\sum_{k\sigma}n_{k\sigma}/S=1-x$, with $S$ the
number of lattice size, as required by the sum rule.\cite{sumrule}
We call the above procedure as a two-mode approach (TMA) for the
excitations in the occupied side (or below the Fermi level). On
the unoccupied side, since $P_G[c^\dagger, P_G]=0$ the
bare-electron and quasi-electron states are identical within the
Hilbert space of the model, and one can prove that the total
spectral weight of $|QE\>$ already satisfies the sum rule
$\sum_{k\sigma}Z_{k\sigma}^+/S=2x$.\cite{Yunoki,sumrule}

We perform Monte Carlo calculations for the energy and spectral
weight of the states $|1\>$ and $|2\>$ below the Fermi level, and
of the state $|QE\>$ above the Fermi level. We also present the
results under SMA for comparison. The calculation is done on a
$10\times10$ lattice. Larger sizes up to $14\times14$ are also
attempted which do not alter the conclusions we reach. In our
calculation we use 200,000 samples to reduce the statistical
error, which we find to be less than 5 meV for quasiparticle
energies. In order to get finer resolution in momentum space we
use four combinations of periodic and anti-periodic boundary
conditions. It turns out that the quasi-hole and quasi-electron
bands are symmetric in energy,\cite{Yunoki} and we define the
central energy as the Fermi level. In Figs.1, the linear size of
the symbols represents the momentum-dependent spectral weight,
while the central position of the symbols represent the
quasi-particle energy as a function of momentum along the cuts
$(0,0)\ra (\pi,\pi)\ra (\pi,0)\ra (0,0)$. The doping levels are
$x=6\%$ (a), $x=20\%$ (b), and $x=30\%$ (c). The filled-blue
circles are for excitations on the unoccupied side, which are
identical under both SMA and TMA. The open-green circles
corresponds to the result given by the SMA, which yields a single
band below the Fermi level with the largest spectral weight at the
$\Gamma$-point. This is a common result under SMA,\cite{Bieri,LiT}
but is inconsistent with the recent ARPES measurements. This
should be contrast to the following TMA results. The filled-red
and filled-pink circles are for excitations on the occupied side
contributed by the lower energy eigenstate $|1\rangle$ and higher
energy eigenstate $|2\rangle$, respectively. We find that along
the nodal direction the spectral weight of $|1\rangle$ state at
the $\Gamma$ point is greatly suppressed and gradually increases
away from the $\Gamma$ point in the nodal direction and antinodal
directions. For instance, in Fig.1(a) $Z_1^{-}(k=0)=0.06$ while
$Z_1^{-}(k\sim k_F)=0.16$. The band width below the Fermi level is
roughly $0.4eV$. In the mean time, we also get a band at higher
binding energy ($1.2\sim1.5$ eV) contributed by the states $|2\>$.
The suppression of low energy spectral weight near the zone center
and the appearance of a higher energy band are exactly what ARPES
reveals. Moreover, the energy scale of the higher energy band is
of order $3t$ rather than the Mott Hubbard gap (which would be
infinite in a t-J model), it therefore reflects the information of
soft lower Hubbard band.\cite{twobandMF}

\begin{figure}
\includegraphics[width=9cm]{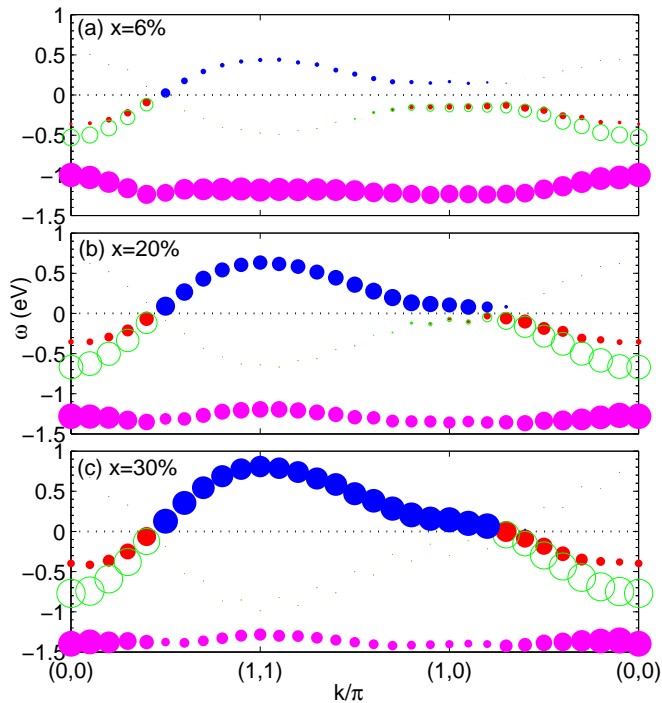}
\caption{(Color online) Energy dispersion and spectral weight map
in the energy-momentum space (along high symmetry cuts). Only
nearest hopping is included in the kinetic part of the
Hamiltonian. The doping level is (a) $x=6\%$, (b) $x=20\%$, (c)
$x=30\%$. The spectral weight scales as linear size of the
symbols. The open (filled) circles are results of SMA (TMA). The
filled-blue, filled-red and filled-pink symbols denote results for
quasi-electron states, 1-band from states $|1\>$ and 2-band from
states $|2\>$, respectively.}
\end{figure}

\begin{figure}
\includegraphics[width=9cm]{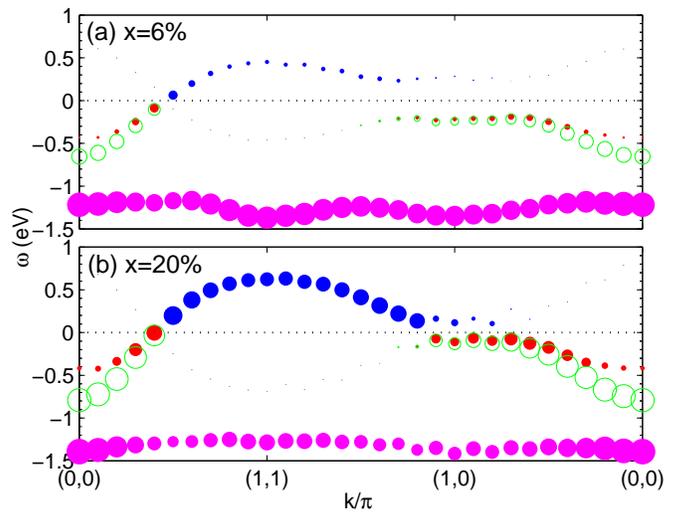}
\caption{(Color online) The same plot as Fig.1(a) except that a
third nearest neighbor hopping is added in the Hamiltonian. }
\end{figure}

By inspecting Figs.1 we see that with increasing doping the
spectral weight near (below or above) the Fermi level increases,
while it decreases on the high energy band below the Fermi level.
This signifies a gradual transfer of the spectral weight from the
high energy band to the low energy bands. On the other hand, the
slope of the dispersion near the nodal Fermi point barely changes
for the 1-band under our TMA. The ratio among the Fermi velocities
we estimated is $1:1.05:1.25$ for $x=6\%$, $20\%$ and $30\%$. This
is rather consistent with the "universal Fermi velocity" observed
by ARPES. (Due to poor screening in cuprates, long-range Coulomb
interaction at lower doping levels may have caused a slightly
larger Fermi velocity on top of the correlation effects under
discussion.) In contrast, the SMA (open-green symbols) gives a
corresponding ratio $1:1.26:1.52$,\cite{Yunoki} and would change
more significantly in the renormalized mean field
theory,\cite{rmft,GrosRMFT} which is rather far from the
experimental result. We notice that the universal Fermi velocity
was also achieved in the literature by tuning $J/t$~\cite{Yunoki}
or by including a three-site hopping term in the
Hamiltonian.\cite{Paramekanti} Finally we observe from Figs.1 that
in the range of doping levels we studied, the energy scales for
the low and high energy bands below the Fermi level does not
change significantly. This is in agreement with the experimental
results.\cite{lanzara,Meevasana}

It turns out that the suppression of the low energy spectral
weight near the zone center can be made more complete by including
the third nearest neighbor hopping integral $t_3$ in the
Hamiltonian and the trial wave function. For doping level $x=6\%$,
$t_1=0.4eV$, $t_3=0.06eV$ and $J=0.12eV$ the optimal variational
parameter in the trial wave function is $t_{var}^{''}=0.12$,
$\Delta_{var}=0.55$, $\mu_{var}=-0.28$. The quasiparticle spectra
are plot in Fig.2(a). In this case $Z_1^{-}(k=0)=0.045$ and
$Z_1^-(k\sim k_F)=0.18$. The curvature of the high energy
dispersion near the zone center is changed as compared to the case
in Fig.1(a). Similar tendency is found in Fig.2(b) as compared to
Fig.1(b). We observe that the curvature also varies by inspecting
published data for different families of
cuprates.\cite{feng,lanzara,Meevasana} We also notice that a
negative second nearest neighbor hopping integral $t_2$ can lead
to a less prominent suppression of the spectral weight near the
zone center, but the existence of a higher energy band is robust.

\begin{figure}
\includegraphics[width=9cm]{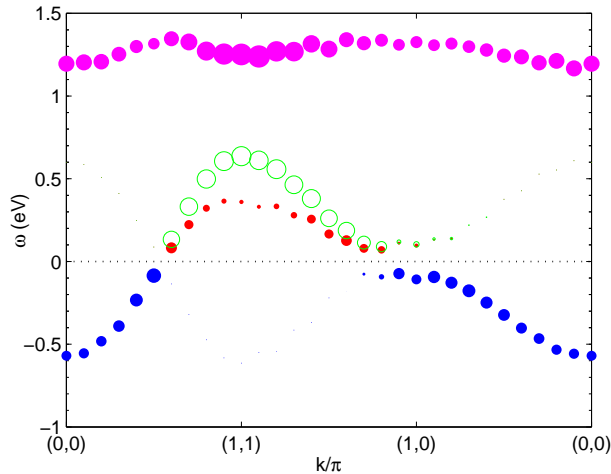}
\caption{(Color online) Energy dispersion and spectral weight map
in the energy-momentum space (along high symmetry cuts) in an
electron doped case with $x=16\%$. Open (filled) symbols show
results of SMA (TMA). The filled-blue, filled-red and filled-pink
symbols denote results for quasi-hole states, 1-band from states
$|1\>$ and 2-band from states $|2\>$ (in the electron picture),
respectively. Notice that the high energy anomaly now appears
above the Fermi level.}
\end{figure}

We have also investigated the quasiparticle excitations in
electron doped cuprates, which is still described by $t-J$ models
after a particle-hole transformation. We choose a doping level
$x=16\%$, at which it is believed that antiferromagnetic order
does not exist. The parameters in the Hamiltonian are $t_1=-0.4eV$
and $J=0.12eV$. The variational parameter are the nearest neighbor
singlet gap function $\Delta_{var}$ with $d$-wave pairing symmetry
and the chemical potential $\mu_{var}$. The optimal parameters we
found are $\Delta_{var}=0.27$ and $\mu_{var}=-0.3$. In order to
relate to the results in ARPES experiments one need to switch back
to the electron picture. Our result is shown in Fig.3. The filled
circles are the results under TMA, and the open circles are from
SMA. We find that in the electron-doped case the quasiparticle
excitations below the Fermi level exhibit neither suppression of
spectral weight at the zone center nor a higher energy band. This
is consistent with the results of exact
diagonalization.\cite{exactD} Instead, a high energy band does
appear but above the Fermi level. In fact, apart from some change
of parameters in the Hamiltonian, the electron-doped dispersion
and the associated spectral weight can be regarded as those of the
hole-doped case but viewed up-side-down.

To conclude, we proposed a two-mode approach to enforce the sum
rules for quasiparticle excitations both below and above the Fermi
level in our Gutzwiller projection variational Monte Carlo study
of doped Mott insulators described by the one-band t-J model. The
TMA resolves a long standing issue regarding the sum rule in
VMC.\cite{Edegger} In the hole doped case, we obtained results in
agreement with the highly unusual features revealed in recent
angle-resolved photoemission spectra: 1) an almost
doping-independent Fermi velocity, 2) two energy scales below the
Fermi level, and 3) suppression of the spectral weight near the
zone center on the low energy band below the Fermi level. In the
electron doped case we predicted that the dispersion below the
Fermi level does not have high energy anomalies.

\acknowledgments{We thank X. G. Wen for enlightening discussions.
This work was supported by NSFC 10325416, the Fok Ying Tung
Education Foundation No.91009, the Ministry of Science and
Technology of China (under the Grant No. 2006CB921802 and
2006CB601002) and the 111 Project (under the Grant No. B07026).
The numerical calculations were performed on the Sun Rack
1000-38.}

\end{document}